\documentclass[conference,a4paper]{APSIPA2025}

\usepackage{amsmath,amsfonts}
\usepackage{graphicx}
\usepackage{threeparttable}
\usepackage[backend=biber,style=ieee,]{biblatex}
\usepackage{multirow}
\usepackage{booktabs}
\usepackage{CJKutf8}
\usepackage{hyperref}
\hypersetup{
    colorlinks=true,
    citecolor=blue,
}
\usepackage{colortbl}
\usepackage{url}
\usepackage{makecell}
\usepackage{cleveref}
\usepackage{bm}

\usepackage{geometry}
\usepackage{fancyhdr}

\fancypagestyle{firststyle}{
  \fancyhf{}
  \fancyhead[C]{2025 Asia Pacific Signal and Information Processing Association Annual Summit and Conference (APSIPA ASC)}
}

\begin{document}

\title{Advancing Speech Quality Assessment Through Scientific Challenges and Open-source Activities}

\author{
\authorblockN{
Wen-Chin Huang
}
\authorblockA{
Nagoya University, Japan
}
}

\maketitle
\thispagestyle{firststyle}
\pagestyle{empty}

\begin{abstract}
Speech quality assessment (SQA) refers to the evaluation of speech quality, and developing an accurate automatic SQA method that reflects human perception has become increasingly important, in order to keep up with the generative AI boom. In recent years, SQA has progressed to a point that researchers started to faithfully use automatic SQA in research papers as a rigorous measurement of goodness for speech generation systems. We believe that the scientific challenges and open-source activities of late have stimulated the growth in this field. In this paper, we review recent challenges as well as open-source implementations and toolkits for SQA, and highlight the importance of maintaining such activities to facilitate the development of not only SQA itself but also generative AI for speech.
\end{abstract}

\section{Introduction}

Evaluating the quality of a speech sample, which is also known as speech quality assessment (SQA) \cite{speech-evaluation-review-2011, sqa-2011, speech-evaluation-review-2024}, can be a complicated process, and human is often considered the gold standard. Not only because the end user is human, but the assessment involves considering multiple dimensions simultaneously, including naturalness, intelligibility, and other intended purposes, which is a highly difficult task. While the main purpose of SQA was to monitor the quality of telecommunication services, its application to evaluate speech generation systems has been gaining attention in recent years due to the generative AI boom. However, having human listeners to judge the speech quality in the development pipeline can be costly, and has thus motivated the development of automated evaluation protocols.

Compared to other attributes like intelligibility, the highly-subjective nature of speech quality \cite{sqa-2011} emphasizes how important it is for a metric to be well-correlated with human perception. It has therefore become increasingly popular to develop SQA methods that are directly optimized using human preference data. Since such approaches are data-driven and thus based on machine learning models, the field of SSQA greatly benefits from the rapid development of deep neural networks (DNNs) in the past decade \cite{automos, mosnet, dnsmos}. As a result, SQA methods nowadays have been shown to correlate well with human ratings, leading to the adaptation of such methods to evaluate speech generation models in scientific papers.

\begin{figure}[t]
    \centering
    \includegraphics[width=\linewidth]{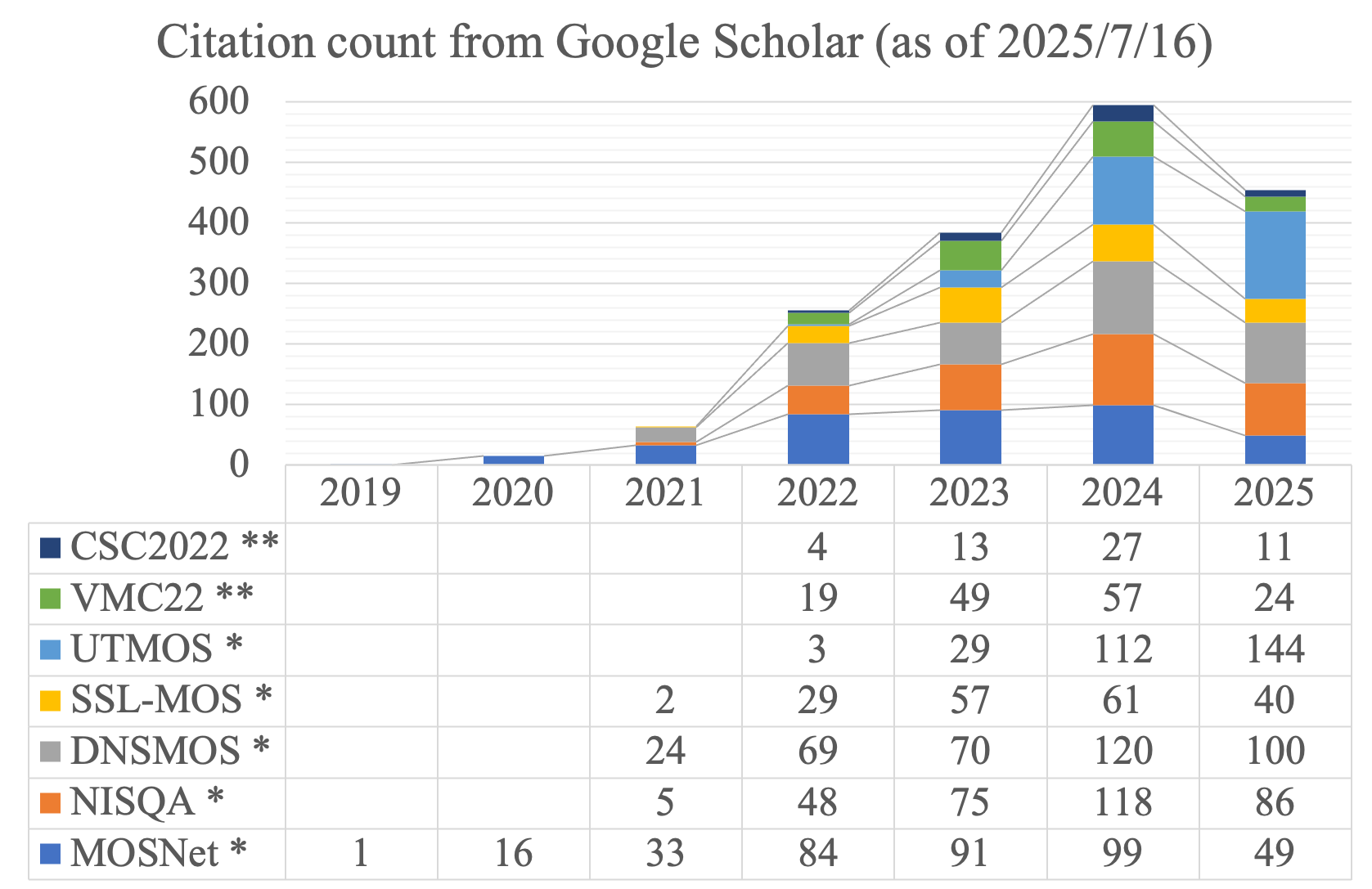}
    \caption{Google Scholar Citations count of recent SQA papers.*: papers on SQA with open-source implementations. **: summary papers of scientific challenges}
    \label{fig:citaitons}
    \vspace{-5mm}
\end{figure}

The success of a research field often arises from the collective efforts of the community as a whole. In the era of deep learning, this can be realized in two key ways. First, scientific challenges can increase visibility and attract interest to a field. These challenges are not about competition or ranking; rather, the main goal is to advance on specific problems. By providing a standardized framework -- including shared datasets and evaluation protocols -- they enable systematic analysis and comparison of different approaches. The insights and findings can further be disseminated to benefit researchers across related fields. As evidence, in 2022, two scientific challenges were organized: the ConferencingSpeech Challenge (CSC) \cite{conferencingspeech2022}, and the VoiceMOS Challenge (VMC) \cite{voicemos2022}. Since then, the citation count of recent SQA papers increased rapidly, as shown in Figure~\ref{fig:citaitons}.

Open-source activities have also contributed to the rapid advancement of SQA. The most direct application of any SQA method is its integration into real-world scenarios. Therefore, ease of use becomes a crucial factor. Conventional non-DNN-based SQA methods such as PESQ \cite{pesq} can be computed analytically. In contrast, DNN-based methods require model training, posing a barrier for users who simply want to assess the quality of, for instance, a speech generation system. Converting human-annotated quality labels into trained models and making them openly accessible enables a broader adoption of modern SQA methods. This, in turn, has played a key role in accelerating progress in the field.

In this perspective paper, we first share our four years of experience in running the VoiceMOS Challenge series, whose scope was further expanded from SQA to quality assessment of general audio, thus rebranded to the AudioMOS Challenge. We share the task design, key insights, and feedback from participants.
Then, we review recent open-source activities in SQA, which greatly benefit the development of this field.
Finally, we discuss future directions.

\begin{table*}[t]

    \scriptsize

    \centering
    \caption{Summary of the tracks in VoiceMOS Challenge 2022-2024 and AudioMOS Challenge 2025.}
    
    \centering
    \begin{tabular}{c c c c c c }
        \toprule
        Challenge & Track & Dataset & Audio type & Evaluation axis & Setting \\ \midrule
            \multirow{2}{*}[-2pt]{VoiceMOS Challenge 2022} & Main & BVCC & English TTS \& VC & Naturalness & in-domain  \\ \cmidrule{2-6}
        & OOD & BC19 & Chinese TTS & Naturalness & in-domain \\
        
        \midrule
            
            \multirow{3}{*}[-5pt]{VoiceMOS Challenge 2023} & Track 1 & BC23 & French TTS & Naturalness & out-of-domain \\ \cmidrule{2-6}
            & Track 2 & SVCC23 & English singing voice conversion & Naturalness & out-of-domain \\ \cmidrule{2-6}
            & Track 3 & TMHINT-QI(S) & Chinese noisy \& enhanced speech & Naturalness & out-of-domain \\
            
            \midrule
            
            \multirow{3}{*}[-18pt]{VoiceMOS Challenge 2024} & Track 1 & Zoomed-in BVCC & English TTS \& VC & Naturalness & \begin{tabular}[c]{@{}c@{}}Same audio,\\zoomed-in labels\end{tabular} \\ \cmidrule{2-6}
            & Track 2 & SingMOS &  \begin{tabular}[c]{@{}c@{}}Chinese \& Japanese\\singing voice synthesis \& conversion\end{tabular} & Naturalness & in-domain \\ \cmidrule{2-6}
            & Track 3 & -- & English noisy \& enhanced speech & \begin{tabular}[c]{@{}c@{}}Signal distortion \&\\background noise \&\\overall quality\end{tabular} & \begin{tabular}[c]{@{}c@{}}out-of-domain,\\semi-supervised\end{tabular} \\
            
            \midrule
            
            \multirow{3}{*}[-20pt]{AudioMOS Challenge 2025} & Track1 & MusicEval & Text-to-music &  \begin{tabular}[c]{@{}c@{}}Overal quality \& \\ Textual alignment\end{tabular} & in-domain \\ \cmidrule{2-6}
            & Track 2 & -- & \begin{tabular}[c]{@{}c@{}}Natural speech/audio/music,\\text-to-speech/audio/music\end{tabular}& \begin{tabular}[c]{@{}c@{}}Product quality \& \\ Product complexity \& \\ Content enjoyment \& \\ Content usefulness\end{tabular} & out-of-domain \\ \cmidrule{2-6}
            & Track 3 & -- & \begin{tabular}[c]{@{}c@{}}English TTS\\at different sampling rate\end{tabular} & Naturalness & \begin{tabular}[c]{@{}c@{}}same audio,\\different listening tests\end{tabular} \\
        
        \bottomrule
    \end{tabular}
    \label{tab:vmc}
\end{table*}

\section{The VoiceMOS and AudioMOS Challenge series}

The VMC series was initiated in 2022, and has been held annually since then\footnote{\url{https://sites.google.com/view/voicemos-challenge}}. The term ``MOS'' stands for ``mean opinion score'', which is a common listening test type \cite{p800}. As the name suggests, the task involves predicting the MOS for a given voice sample. From 2022 to 2024, the challenge focused primarily on the evaluation of speech. In 2025, however, the scope was broadened to include other audio modalities such as music and general environmental sounds. To reflect this expansion, the challenge was rebranded as the AudioMOS Challenge. Table~\ref{tab:vmc} summarizes the tracks in each year.

In SQA, out-of-domain generalization is a crucial aspect, due to the nature of how listening tests are conducted: each listening test represents a unique context, with different contents (text, speakers, etc.), recruited listeners, ranges of systems being evaluated, and even instructions. Thus, with respect to an SQA model trained on a specific dataset, the testing scenario can be either \textit{in-domain} or \textit{out-of-domain}, where the former means the test samples and the ratings come from the same listening test as that of the training set, and the latter means they come from different listening tests.

The challenge was hosted on CodaLab \cite{codalab} (which was further renamed to CodaBench\footnote{\url{https://www.codabench.org/}}). 
In each year, the challenge was divided into the training phase, evaluation phase, and post-challenge. The training phase was usually two to three months long, and participants could use the training set (if provided) to develop their system. In the evaluation phase, which was always one week long, participants made their predictions of the quality of the test set samples. After the submission deadline, we released the test set ground truth labels for participants to perform analysis and wrap up their paper. We also asked each team to submit a system description form based on the template we distributed, including surveys on their opinions on the challenge and future directions.

To assess SQA models, several evaluation criteria were used, but for most of the time, we used system-level Spearman rank correlation coefficient (SRCC) as the primary metric for determining the ranking.

Each year, we provided one or more open-sourced baseline systems, giving participants access to pretrained models along with recipes for training, fine-tuning, and making predictions on the challenge datasets. 
This component is essential to a scientific challenge for several reasons. First, it significantly lowers the barrier to participation. Second, the baseline system typically represents the state-of-the-art or the most representative approach available at the time. As a result, it becomes feasible to assess whether the challenge drives meaningful progress in the field.

\subsection{The VoiceMOS Challenge 2022}

\subsubsection{Track and setting}

There were two tracks in VMC 2022, namely the main track and the out-of-domain track. The main track was based on a large-scale dataset of MOS ratings for synthesized audio samples as well as reference natural speech samples, which we collected in 2021 \cite{bvcc}. We mainly collected English-language synthesized audio samples from several past Blizzard Challenges (BCs) from 2008-2016 and from all previous Voice Conversion Challenges, as well as publicly-shared samples from ESPnet-TTS, one of the most commonly used TTS toolkits at that time \cite{espnet}. Altogether, we collected samples from 187 different systems, and we conducted a large-scale listening test where each sample received eight ratings. Although we carefully designed a training, development, and testing split of the data while holding out some unseen speakers, synthesis systems, and listeners in the development and test sets, we still consider this track to be an ``in-domain'' setting.

Additionally, we also ran an out-of-domain (OOD) track by making use of the BC 2019 data, which focused on Chinese text-to-speech (TTS) samples, and provided participants with a very small amount of labeled training data (136 samples). Although we also designed the training, development, and testing split to include unseen systems and listeners, since labeled training data was provided, we considered the OOD track to be somewhat ``in-domain''. Still, this track was challenging both in terms of the smaller amount of labeled training data, as well as the language mismatch with respect to the main track.

\subsubsection{Results and insights}

The challenge attracted \textbf{22} participating teams from academia and industry, which we consider to be very successful compared to popular scientific challenges like the ImageNet Large Scale Visual Recognition Challenge \cite{ilsvrc}, which had on average 20-30 participants per year. In addition, we provided three baselines \cite{ssl-mos, ldnet, mosanet}. For the main track, the best baseline had a system-level MSE and SRCC of 0.148 and 0.921, ranking 18th and 12th in the two metrics, respectively. The top prediction systems in system-level MSE and SRCC scored \textbf{0.090} and \textbf{0.939}, respectively. As for the OOD track, the top prediction systems had an MSE of \textbf{0.030} and a SRCC of \textbf{0.979}, respectively. 

Overall, we observed that fine-tuning SSL models for the MOS prediction task was a powerful approach that can produce predictions with \textbf{very high correlations with real listener ratings}, even in the case of the OOD track where a very small amount of label was available. Feedback from participants was about including a larger variety of audio to evaluate, including synthesis in more different languages, singing voice synthesis, and noisy and enhanced speech.

\subsection{The VoiceMOS Challenge 2023}

The outcomes of the first challenge motivated our design of the 2023 edition of the challenge \cite{voicemos2023}. If, even with a very small amount of labeled training data, the correlation coefficient between the machine's prediction and that of humans can be larger than 0.95, how about an entirely \textbf{out-of-domain, zero-shot} setting? Therefore, we focused on real-life MOS prediction in a variety of speech domains, as the labels in the test sets are completely new listening tests: they were collected at the same time as the challenge, meaning that team predictions were made before the actual ground-truth MOS values were known to anyone.  

\subsubsection{Data and tracks}

There were three tracks in VMC 2023. The first track was based on the Blizzard Challenge 2023 \cite{bc2023}, which focused on French text-to-speech synthesis. The second track was based on the Singing Voice Conversion Challenge 2023 (SVCC) \cite{svcc2023}, where the input singing voice sample was converted to a different speaker identity, using either sung (matched) or spoken (mismatched) reference audio from the target speaker. Unlike VMC 2022, where the listening tests were completed in advance and divided into training, development, and test sets, when the VMC 2023 training phase took place, the Blizzard and SVCC listening tests were still ongoing. Thus, no official training data was provided for these two tracks.

For the third track, we considered the quality assessment of noisy and enhanced speech for the first time. Unlike the previous two tracks, we provided the TMHINT-QI \cite{tmhintqi} dataset as training data, and for the test set, a separate listening test was conducted, with the same noise generation process, partially different speech enhancement systems, and completely different raters. We named the test set TMHINT-QI(S) \cite{tmhintqis}. The design of this track was to answer the substantial interest from the participating teams in VMC 2022, as we also noticed many parallel efforts towards more automatic evaluation methodologies in the speech synthesis and speech enhancement communities. Considering that these are similar tasks, we believed there could be benefits from more communication and collaboration between these communities.

\subsubsection{Results and insights}

In total, \textbf{ten} teams participated in the 2023 challenge. This year we also provided two baseline systems: one was SSL-MOS, the best baseline system in VMC 2022 \cite{ssl-mos}, and the other one was UTMOS, the best performing system in VMC 2022 \cite{utmos}. In each track, at least one team outperformed the baseline, showing that progress was indeed made. The most important result was that most teams' scores for the different tracks are very different, and \textbf{no team had high scores on all tracks using the same model trained on the same data}, indicating that \textbf{general-purpose MOS prediction can still be considered an open research problem}.  

As for other findings, we were surprised that many teams performed well on the second track, whose focus was singing voice conversion samples. At the time being, quality assessment for singing voices was still an underexplored field, so we suspected that the domain mismatch between synthesized singing and speech was not as large as we had assumed. Looking at each team's approach, we found that listener-dependent modeling \cite{ldnet} was more popular this year, and teams that used a mix of different training datasets also tended to do better.


\subsection{The VoiceMOS Challenge 2024}

\subsubsection{Data and tracks}

There were three tracks in VMC 2024 \cite{voicemos2024}. The first track was MOS prediction for ``zoomed-in'' systems, motivated by a real-world scenario where the researchers wish to evaluate a speech generation model under development, whose quality is expected to be better than any previous system. Based on the BVCC dataset, we collected a set of ``zoomed-in'' subjective ratings \cite{range-equalizing-bias}, where new listening tests were conducted using approximately 50\%, 25\%, 12\%, and 6\% of the highest-rated systems. No new training data was provided, and the test set consists of 1000 samples from the 25\% subset, 500 of which are also included in the 12\% subset. Although the samples in the test sets are all from BVCC, since new listening tests were conducted, this track was considered out-of-domain.

The second track was based on SingMOS \cite{singmos}, a newly collected dataset consisting of natural singing voice samples, vocoder analysis-synthesis samples, and singing voice synthesis/conversion samples in Chinese and Japanese. The official split was used, with partially unseen samples in the training set. Thus, this track was considered in-domain.

The third track focused on MOS prediction for noisy and enhanced speech. While this may appear similar to the VMC 2023 track 3, we introduced two key novelties. First, instead of relying on a single MOS value, the evaluation employed three perceptual dimensions defined in ITU-T P.835 \cite{p835}: signal quality (SIG), background intrusiveness (BAK), and overall quality (OVRL). Second, the training data was further limited in size. Specifically, the training and validation sets were derived from the UDASE task of the 7th CHiME Challenge \cite{chime7-udase, chime7-evaluation}, containing only 60 and 40 samples, respectively. The evaluation set was constructed using data from a separate listening test involving samples from the VoiceBank-DEMAND dataset \cite{Voicebank_Demand}.

\subsubsection{Results and insights}

This year, we received submissions from five teams from academia and three teams from industry, for \textbf{eight} in total from six different countries. We also had a baseline system for each track. For track 1, although multiple teams outperformed the baseline, the overall performance was significantly lower than the original labels, highlighting the difficulty in ranking high-quality speech synthesis systems. For track 2, although no team outperformed the baseline in terms of system-level SRCC, all systems had a system-level SRCC higher than 0.8. While participants questioned whether the baseline was too strong, we suspect that the overly-easy in-domain setting was the main cause. Finally, in track 3, among the three axes, SIG was the most difficult dimension to predict, and participating teams had diverse behaviors.

\subsection{The AudioMOS Challenge 2025}

The rapid progress of music and general audio generation led to an urgent need for an automatic evaluation method for text-to-music (TTM) and text-to-audio (TTA) systems that reflect human perception, as demanded by participants in the 2024 challenge. In light of this, we expanded the VMC series and rebranded to the AudioMOS Challenge (AMC).

\subsubsection{Data and tracks}

There were three tracks in AMC 2025. The first track focused on MOS prediction of TTM systems, where we used MusicEval \cite{musiceval}, a dataset containing music clips generated by 31 modern TTM systems. Music experts were recruited to rate each clip in terms of overall musical quality and alignment with the text prompt, which respectively emphasizes the importance of both the quality of the generated music and its consistency with the given prompt.
The second track was based on Meta Audiobox Aesthetics \cite{aes}, a suite of unified assessment methods for speech, music, and sound. Instead of a single MOS, the evaluation protocol consists of four new evaluation dimensions: production quality, production complexity, content enjoyment, and content usefulness. The task was to assess synthetic samples from TTS, TTA, and TTM along the four axes.
The third track focused on MOS prediction for synthesized speech in different sampling frequencies.  During the training phase, participants were provided with samples in 16, 24, and 48 kHz, along with their ratings obtained from listening tests that only contained samples in the same sampling frequencies. For the test set, the participants were asked to make predictions of synthetic samples that reflect their scores in a listening test that contains samples from all sampling frequencies.

\subsubsection{Results and insights}

We received submissions from \textbf{24} unique teams, which was the most among the previous VMCs, demonstrating the increasing interest in audio quality assessment. We also prepared a baseline for each track. For track 1, not only did all teams outperform the baseline, but the best-performing team achieved system-level SRCCs over 0.95 on both axes, again reflecting the overly-easy in-domain setting. For the second track, as it was difficult for a single system to excel in the prediction of all four axes, more than half of the teams outperformed the baseline. For track 3, although all teams outperformed the baseline, the limited number of synthesis systems in the dataset made many participants question whether the track was properly designed.

\subsection{Key factors to the success of a challenge}

The most important aspect of organizing a scientific challenge is attracting a sufficient number of participants. Across the four editions, we observed fluctuating levels of participation, with 22, 10, 8, and 24 teams joining in 2022 through 2025, respectively. Upon reflection, we identified several key factors that contributed to higher engagement:

\begin{itemize}
    \item \textbf{Well-defined task}: Although tasks such as the “zoomed-in” setting and the “different sampling rate” condition were frequently requested by past participants, it proved challenging for the organizers to design task setups that were both fair and easy to understand. Striking a balance between research novelty and clarity of formulation remains a key difficulty in challenge design.
    \item \textbf{User-friendly baseline}: In VMC 2022 and AMC 2025, we dedicated more effort to developing comprehensive and easy-to-use baseline implementations, including pretrained models, clear documentation, and ready-to-run scripts. These efforts greatly lowered the entry barrier for new participants, especially those unfamiliar with the task.
    \item \textbf{Marketing and advertisement}: This is arguably the most critical yet often underestimated factor in organizing a successful challenge. In 2023 and 2024, we did not actively promote the challenge through mailing lists and social media, and this lack of visibility likely contributed to the lower participation in those years.
\end{itemize}

\begin{table}[t]
    \centering
    \caption{Comparison of existing open-sourced speech quality assessment methods.}
    \label{tab:toolkits}
    
    \begin{tabular}{c | c c c c}
        \toprule
        Name & Inference &  \makecell{Model\\training} & \makecell{Multi-\\model} & \makecell{Multi-\\dataset} \\ 
        \midrule
        MOSNet \cite{mosnet} & \checkmark & \checkmark & & \\
        DNSMOS \cite{dnsmos} & \checkmark & & \checkmark & \\
        NISQA \cite{nisqa} & \checkmark & \checkmark & & \\
        SSL-MOS \cite{ssl-mos} & \checkmark & \checkmark & & \\
        UTMOS \cite{utmos} & \checkmark & \checkmark & & \\
        TorchAudio-Squim \cite{torchaudio-squim} & \checkmark & & \checkmark & \\
        SHEET \cite{sheet} & \checkmark & \checkmark & \checkmark & \checkmark \\
        VERSA \cite{versa} & \checkmark & & \checkmark & \\
        \bottomrule
    \end{tabular}
    \vspace{-5mm}
\end{table}

\section{Open-source toolkits}

In this section, we are particularly interested in SQA methods based on DNNs. This means they cannot be expressed in analytical forms, and are too costly for users to train the models from scratch. As a result, open-sourcing these models and making them easy to use becomes very important.

Table~\ref{tab:toolkits} shows a comparison between representative open-source SQA methods. In the early stages of SQA development, many publicly available codebases were primarily released as supplementary materials to their corresponding scientific papers, including MOSNet, DNSMOS, NISQA, SSL-MOS, and UTMOS \cite{mosnet, dnsmos, nisqa, ssl-mos, utmos}. While open-sourcing code has become a standard expectation in modern machine learning research, it was relatively uncommon in a smaller field like SQA at the time. This scarcity contributed to the widespread adoption of these early methods. However, these toolkits typically supported only the specific method proposed in the original paper, and were often trained on a single dataset, limiting their generalizability and practical applicability.

To support the \textit{research} development of SQA systems, the SHEET toolkit \cite{sheet} was developed with a particular focus on providing complete \textit{training} and evaluation scripts, supporting a large collection of datasets and several representative SQA models. The goal was to provide a benchmark playground for both experienced researchers and newcomers to easily start working on this area. 

Another line of work aims to provide a user-friendly interface to multiple off-the-shelf metrics and pre-trained SQA models, with the goal to enhance the accessibility of existing SQA metrics. For instance, TorchAudio-Squim \cite{torchaudio-squim} was designed to provide non-intrusive, reference-free quality measures for speech. Although only providing four metrics, its tight integration with TorchAudio \cite{torchaudio}, the official audio domain library of PyTorch, made it an easy-to-use building block for speech signal processing system development. Recently, the VERSA toolkit \cite{versa} was introduced as a unified, lightweight evaluation toolkit designed for not only speech but music and general audio. VERSA was already supporting 65 metrics with 729 variations based on different configurations by the time the paper was published, and it is planned to be continuously developed.

\section{Conclusion and discussions}

In this perspective paper, we first reflected on our experiences organizing the VoiceMOS and AudioMOS Challenge series over the past four years, highlighting key lessons on task design, baseline development, and community engagement. We also reviewed recent progress in open-source SQA toolkits and their role in accelerating research. These collective efforts have definitely inspired innovation and collaboration.

As challenge organizers, we have been constantly listening to the voice of the community. Although the field continues to expand toward general audio quality assessment, we would like to point out that despite the progress, automatic quality assessment of speech itself is still not solved, as we still constantly receive requests on evaluating more complex speech types like multi-lingual speech, expressive TTS, and prompt-based TTS. In addition, we are also seeing benchmarks that are publicly available and continuously evolving \cite{mos-bench, ttsds2}. We hope this paper not only serves as a reflection on past efforts but also as a call to continue building a shared foundation for the future of SQA and beyond.

\noindent\textbf{Acknowledgments:} This work was partly supported by JSPS KAKENHI Grant Number 25K00143.

\printbibliography

\end{document}